\documentclass[aps,prd,preprint]{revtex4}
\usepackage{amssymb,amsmath}
\usepackage{graphicx}
\usepackage{epstopdf}
\begin{document}
\title{Bayesian Reweighting for Global Fits}
\author{Nobuo Sato, J.~F.~Owens, and Harrison Prosper}
\affiliation{Department of Physics, Florida State University, Tallahassee, 
Florida 32306-4350}


\date{\today}

\begin{abstract}

Two different techniques for adding additional data sets to existing 
global fits using Bayesian reweighting have been proposed in the literature. 
The derivation of each reweighting formalism is critically reviewed. A simple 
example is constructed that conclusively favors one of the two 
formalisms. The effects of this choice for global fits is discussed.

\end{abstract}

\maketitle

\section{Introduction}
In general, the understanding of a given phenomenon relies on our ability to
construct a model that describes the relevant data and their corresponding 
uncertainties. 
One way to summarize what the data tell us about a model is to find a
probability 
density function for its parameters.
For such a task, standard fitting techniques such as $\chi^2$ minimization
are commonly used to determine this probability density.  
Typically, this process is iterative: once new data are available, 
a new fit is performed combining the old and new data.
We shall refer to this procedure as a \emph{global fit}.
In some cases, the complexity of the model is such that its numerical 
evaluation makes the fitting procedure time consuming.
For practical reasons, it would be desirable to update the probability 
density by incorporating the information from new data without having to 
perform a full global fit.
Such updating can be achieved by a statistical inference procedure, based on 
Bayes theorem, known as the reweighting technique. 

A particular example, where the reweighting technique is useful, 
is in the context of global fits for the determination of parton distribution 
functions (PDFs).
Modeling and fitting these functions has been the central task of 
several collaborations, e.g., CTEQ, CJ, MSTW, and NNPDF, among others. 
But still, there are kinematic regions where the PDFs are relatively 
unconstrained.
Given the complexity of the calculations,  it is desirable to use 
the reweighting technique to update our knowledge of the PDFs or to quantify 
the potential impact of anticipated data sets on the PDFs.

The idea of reweighting PDFs was originally proposed in~\cite{Giele:1998gw}
and later discussed by the NNPDF collaboration 
in~\cite{Ball:2010gb,Ball:2011gg}.
However, there is disagreement about the reweighting 
procedure, which has led to methods that  differ mathematically.
The purpose of this paper is to discuss the differences between the 
reweighting methods. In particular, we investigate the degree to which 
the reweighting 
procedures yield results that are consistent with those from global fits. 
We shall argue that this
is the case for the method proposed in~\cite{Giele:1998gw}.
 
The paper is organized as follows. In Sec.~\ref{sec:the reweighting method},
 we describe the basics of the reweighting technique. 
In Sec.~\ref{sec:the NNPDF paradox}, we will discuss subtleties in the NNPDF
arguments.
In Sec.~\ref{sec:numerical example}, we will present a simple 
numerical example to display the differences between the reweighting methods.
Our conclusions are given in Sec.~\ref{sec:conclusions}. 
\section{The Reweighting Method}
\label{sec:the reweighting method}
The reweighting of probability densities in order to incorporate the
information from new data is merely the recursive application of Bayes 
theorem. 
Suppose a probability density function (pdf) $\mathcal{P}(\vec{\alpha})$ of 
the parameters 
$\vec{\alpha}$ in a model is known. (To avoid confusion, we shall take 
``PDF" to mean parton distribution function,
 and ``pdf" to mean probability density function.)
Given new data $D$, Bayes theorem states that
\begin{align}
\mathcal{P}(\vec{\alpha}|D)
=\frac{\mathcal{P}(D|\vec{\alpha})}{\mathcal{P}(D)} \mathcal{P}(\vec{\alpha}),
\label{eq:bayes}
\end{align}
where $\mathcal{P}(\vec{\alpha}|D)$, known as \emph{posterior} density, 
is the updated pdf from the \emph{prior} density (or prior for short) 
$\mathcal{P}(\vec{\alpha})$, which can serve as the prior in a subsequent 
analysis. 
The quantity $\mathcal{P}(D|\vec{\alpha})$ called the \emph{likelihood} 
function,
represents the conditional probability for a data set $D$ given the
parameters $\vec{\alpha}$ of the model.
The quantity $\mathcal{P}(D) $ ensures the normalization of the 
posterior density.  
With the new data, the expectation value of an observable 
$\mathcal{O}$ can be written as,
\begin{align}
\text{E}[\mathcal{O}]
&=	\int d^n\alpha \mathcal{P}(\vec{\alpha}|D)
	\mathcal{O}(\vec{\alpha})\notag\\
&=	\int d^n\alpha \frac{\mathcal{P}(D|\vec{\alpha})}{\mathcal{P}(D)} 
	\mathcal{P}(\vec{\alpha})\mathcal{O}(\vec{\alpha})\notag\\
&=	\frac{1}{N} \sum_k w_k \mathcal{O}(\vec{\alpha}_k).
\label{eq:E}
\end{align}
In the last line, we have used a Monte Carlo approximation of the integral in 
which
the parameters $\{\vec{\alpha}_k\}$ are distributed according to the prior
$\mathcal{P}(\vec{\alpha}_k)$.
Similarly, the variance is given by
\begin{align}
\text{Var}[\mathcal{O}]
&=\frac{1}{N} \sum_k w_k (\mathcal{O}(\vec{\alpha}_k)-\text{E}
[\mathcal{O}])^2 .
\label{eq:Var}
\end{align}
The quantities $\{w_k\}$ are \emph{weights} that are proportional to 
$\mathcal{P}(D|\vec{\alpha}_k)$. 
Their normalization is fixed by demanding $\text{E}[1]=1$, that is,  
$\sum_k w_k=N$.

The reweighting procedure depends on the form assumed for the likelihood 
function.
The form of the likelihood function is not unique since it depends on the 
amount of information 
we want to extract from the new data. 
To clarify, suppose the new data consist of $n$ data points $\{(x_i,y_i)\}$
with uncertainties in $\{y_i\}$ given by a covariance matrix $\Sigma$.
Let us call $\{t_i=f(x_i,\vec{\alpha})\}$ the $n$ predictions from the model 
$f$ with parameters $\vec{\alpha}$.
Assuming a Gaussian model, the conditional probability for new data to
be confined in a differential volume $d^ny$ around $\vec{y}$ for a given 
configuration of parameters $\vec{\alpha}$ is
\begin{align}
\mathcal{P}(\vec{y}|\vec{\alpha}) \, d^n y = 
\frac{1}{(2\pi)^{n/2} |\Sigma|^{1/2}} 
e^{-\frac{1}{2} \chi^2(\vec{y},\vec{t}\,)} \, d^n y,
\label{eq:lh1}
\end{align}
where the $\chi^2(\vec{y},\vec{t})$  is defined in the standard way
\begin{align}
\chi^2(\vec{y},\vec{t})
=(\vec{y}-\vec{t})^t \,\Sigma^{-1} \,(\vec{y}-\vec{t}).
\end{align}
On the other hand, we might be interested in the probability for the 
new data to be confined only in a differential shell $\chi$ to $\chi+d\chi$.
This probability density can be obtained by integrating 
$\mathcal{P}(y|\vec{\alpha})$ inside the shell 
(see Appendix \ref{sec:lh2}).
The result,
\begin{align}
\mathcal{P}(\chi|\vec{\alpha})\,d\chi
= \frac{1}{2^{n/2-1} \Gamma(n/2)}\chi^{n-1}e^{-\frac{1}{2} \chi^2} \, d\chi,
\label{eq:lh2}
\end{align}
is the well-known $\chi^2$ distribution.
Using the functions from Eqs.~(\ref{eq:lh1}) and (\ref{eq:lh2}) as likelihoods 
in Eq.~(\ref{eq:bayes}), we obtain the corresponding posterior densities
and weights,
\begin{align}
\mathcal{P}(\vec{\alpha}|\vec{y})
=	\frac{\mathcal{P}(\vec{y}|\vec{\alpha})}{\mathcal{P}(\vec{y})} 
	\mathcal{P}(\vec{\alpha})
\quad\rightarrow\quad
	w_k \propto \exp\left(\frac{1}{2}\chi^2(\vec{y},\vec{t}_k)\right),
\label{eq:bayes1}
\end{align}
\begin{align}
\mathcal{P}(\vec{\alpha}|\chi)
=	\frac{\mathcal{P}(\chi|\vec{\alpha})}{\mathcal{P}(\chi)} 
	\mathcal{P}(\vec{\alpha})
\quad\rightarrow\quad
	w_k \propto \left(\chi^2 (\vec{y},\vec{t}_k)\right)^{\frac{1}{2}(n-1)}
	\exp\left(\frac{1}{2}\chi^2(\vec{y},\vec{t}_k)\right).
\label{eq:bayes2}
\end{align}
Note that $\mathcal{P}(\chi|\vec{\alpha})$ has less information than 
$\mathcal{P}(\vec{y}|\vec{\alpha})$: a given data set $y$ uniquely determines
$\chi$, but a given $\chi$ is consistent with infinitely many data sets $y$.
Therefore, the posterior density $\mathcal{P}(\vec{\alpha}|\chi)$ has 
less information than $\mathcal{P}(\vec{\alpha}|\vec{y})$, a mathematical fact
that we shall quantify using a standard information-theoretic measure.

\section{The NNPDF Paradox}
\label{sec:the NNPDF paradox}
The NNPDF collaboration argues in Ref.~\cite{Ball:2010gb,Ball:2011gg}
that we should avoid 
the use of the likelihood $\mathcal{P}(y|\vec{\alpha)}$ because of the 
Borel-Kolmogorov paradox,
the observation that conditional probabilities, such as 
$\text{Pr}(\vec{\alpha},y=D)/\text{Pr}(y=D)$, for continuous variables $y$ 
are ambiguous because they condition on a set of measure zero for which 
the probability is strictly zero. In the present context, the probability 
that $y$ is \emph{exactly} equal to $D$ is zero. 
In order to give meaning to $\text{Pr}(f, y=D) / \text{Pr}(y = D)$, 
the latter must be defined by a limit. 
The paradox arises because different ways of taking the limit can yield 
different results. 
However, no issue arises in Bayes theorem, Eq.~(\ref{eq:bayes}), even 
when $y = D$
is multidimensional. 
Indeed, statisticians (and some physicists) routinely use multivariate 
densities in Bayes theorem. 
Our intuitive understanding of why Eq.~(\ref{eq:bayes}) is mathematically 
sound is that, first, the probabilities are defined by integrals about 
the point $y$ and second that the \emph{shapes} of the sequence of nested 
sets about the limit point $y$ is the same in both the numerator and the 
denominator. 
When the sets become sufficiently small, the integrals can be approximated 
by the probability density times a small, but \emph{finite}, volume element 
which cancels in Bayes theorem. 
Crucially, this is true for \emph{all} shapes of the $n$-dimensional, small 
but \emph{finite}, volume element and therefore for all sequences of 
(measurable) sets.  
Therefore, the suggestion by the NNPDF authors that, in effect, Bayes theorem, 
Eq.~(\ref{eq:bayes}), is problematic when $y = D$ is multidimensional is
not convincing. 
Indeed, Eq.~(\ref{eq:bayes}) is the bedrock of state-of-the-art Bayesian 
analyses (see for example, Ref.~\cite{bib:Bernardo}.)

\section{A simple numerical example}
\label{sec:numerical example}
This section aims to study the differences between the reweighting results 
from the likelihoods $\mathcal{P}(\vec{y}|\vec{\alpha}) $
and $\mathcal{P}(\chi|\vec{\alpha})$ by a numerical example. 
Simulated data from the function 
\begin{align} 
f(x,\vec{\alpha})=x^{\alpha_0}(1-x)^{\alpha_1},
\label{eq:func}
\end{align}
are generated by adding gaussian noise with independent random variances 
for each value of $x$.
The parameters of the function has been arbitrary set to 
$\vec{\alpha}=(-2,2)$ and a sample 100 points equally spaced in the 
range $0<x<1$ is taken.
This particular functional form is inspired by a typical parton 
distribution function parametrization used in global fits.

For the analysis, the data are divided  into 11 equally spaced regions 
in $x$ and labeled as $\{d_0,d_1,...,d_{10}\}$ from the lowest $x$ region
($d_0$) to the highest $x$ region ($d_{10}$).
Then, the data sets are organized as described in table \ref{tab:datasplit}.
Using $\chi^2$ minimization, we perform global fits to each data set $A_i$.
The uncertainties in the fitted parameters are obtained using the Hessian 
method (see Appendix \ref{sec:hessian}).
As a result we obtain  four parameter vectors  
$\vec{\alpha}_j^{\pm}=\vec{\alpha}_0\pm \delta \vec{\alpha}_j$ 
with $j=1,2$ for each data set $A_i$.
These vectors encode the $1\sigma$ confidence interval of the fitted 
parameters.
For the reweighting, is necessary to construct a Monte Carlo representation 
of the fitted results.
This is done by sampling the parameters as 
\begin{align}
\vec{\alpha}_k=\vec{\alpha_0}+\sum_j \delta\vec{\alpha}_j R_{kj}
\end{align}
where $R_{kj}$ are normally distributed random numbers with variance 1 
and mean 0. $k$ is the number of samples.
Evaluating Eq.~(\ref{eq:func}) with parameters $\vec{\alpha_k}$ from the 
fit $A_i$ yields the desired Monte Carlo sample $\{f_k|A_i\}$.
We compute the latter and its corresponding expectation value 
$\text{E}[f|A_i]$ and variance $\text{Var}[f|A_i]$ for each set $A_i$.
In order to perform the reweighting, we select the Monte Carlo sample 
$\{f_k|A_0\}$ as the \emph{prior} to be reweighted.
Using the data sets $\{B_i\}$ as new evidence, we compute the expectation 
value $\text{E}[f|A_0,B_i]$ and variance $\text{Var}[f|A_0,B_i]$ 
using Eqs.~(\ref{eq:E})and (\ref{eq:Var}) with the weights from 
Eq.~(\ref{eq:bayes1}) or Eq.~(\ref{eq:bayes2}) for each set $B_i$.

\begin{figure}[h!]
\centering
\includegraphics[width=0.85\textwidth]{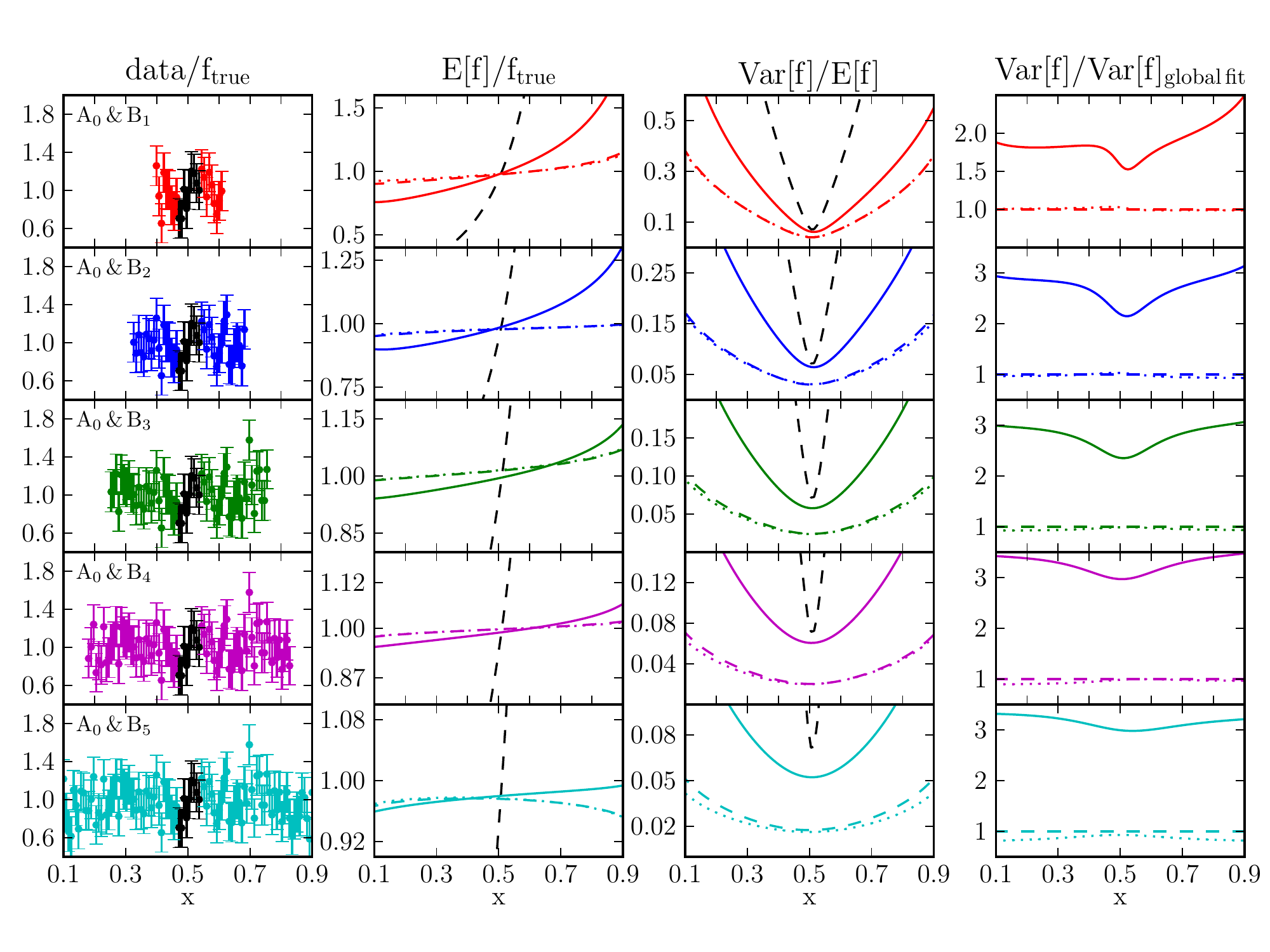}
\caption{
Column 1 shows the data $A_0$ (black) from which the prior distribution 
is obtained and the new evidence $B_i$ (colored) that is used for 
reweighting or appended to $A_0$ to perform a global fit. 
The data is normalized respect to the ``true'' model.
Columns 2,3,4 shows expectation values and variances from global fits and
reweighting.
Dashed lines are the results from global fits. 
Black dashed uses only the data $A_0$ while the colored dashed line includes
the new evidences.
Solid and dotted lines are reweighting results of data set $A_0$ using
the evidences of data $B_i$.
Dotted uses $w_k\propto \exp{(-\frac{1}{2}\chi^2_k)}$ while solid 
uses $w_k\propto \chi_k^{(n-1)}\exp{(-\frac{1}{2}\chi^2_k)}$. 
}
\label{fig:example5A}
\end{figure}

The results are shown in Fig.~\ref{fig:example5A} where a clear disagreement 
between the two reweighting methods is exhibited. 
The variances obtained by using the likelihood 
$\mathcal{P}(\chi|\vec{\alpha})$ are greater than the variances obtained 
from the likelihood $\mathcal{P}(\vec{y}|\vec{\alpha})$ and the convergence 
of the expectation values is much faster for the latter case.
This is consistent with the discussion in section 
\ref{sec:the reweighting method} where we argued that the posterior 
$\mathcal{P}(\vec{\alpha}|\chi)$ contains less information than
$\mathcal{P}(\vec{\alpha}|\vec{y})$. 
More importantly, reweighting with the likelihood
$\mathcal{P}(\vec{y}|\vec{\alpha})$ yields a  
result that is more compatible with that obtained from the global fits than
is that obtained using the likelihood $\mathcal{P}(\chi|\vec{\alpha})$. This 
is illustrated by the dotted and dashed curves being nearly identical while 
the solid and dashed curves show significant differences.

In the light of above, it is important to discuss why the NNPDF collaboration
has obtained reweighting results compatible with global fits 
in~\cite{Ball:2010gb,Ball:2011gg}
even when they have used the likelihood  
$\mathcal{P}(\chi|\vec{\alpha})$ instead of
$\mathcal{P}(\vec{y}|\vec{\alpha})$. 
In their case, their prior corresponds to PDFs fitted using 
deep inelastic scattering data (DIS) and Lepton Pair Production data (LPP).
By performing the reweighting and comparing it with a new global fit 
using the W-lepton asymmetry data, they have proven the consistency 
of their reweighting method.
However, it is also known that PDFs are already reasonably well constrained by 
the DIS and LPP data.
This means that the information provided by the W-lepton data is sub-dominant
with respect to the DIS and LPP data. 

\begin{figure}[h!]
\centering
\includegraphics[width=0.85\textwidth]{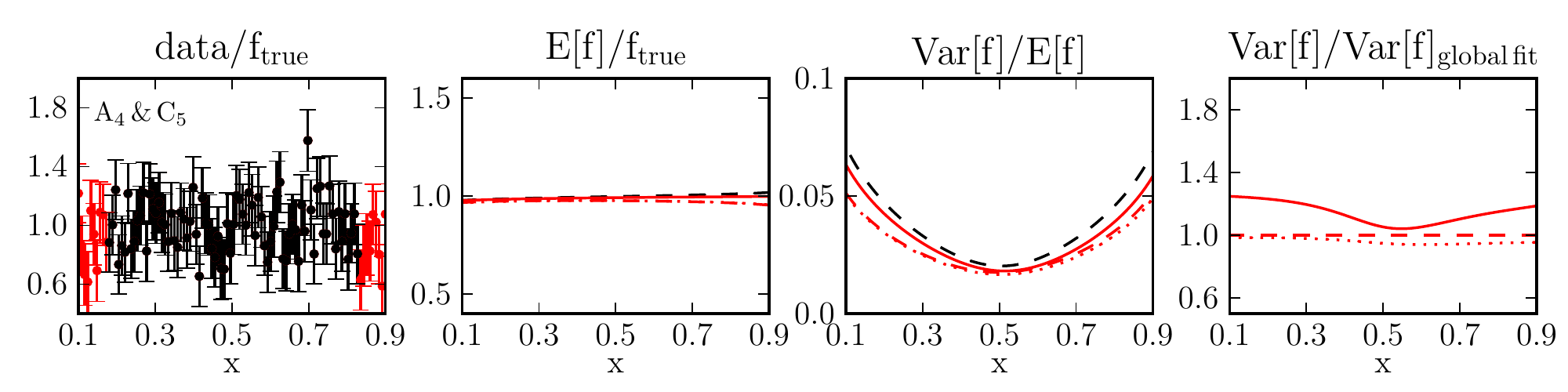}
\caption{
Similar to Fig.~\ref{fig:example5B}. In this case, set $A_4$ is used 
to obtain the prior distribution and set $C_5$ is used for reweighting
or appended to $A_4$ for a global fit.
}
\label{fig:example5B}
\end{figure}

We have performed a similar exercise as before but this time using 
the Monte Carlo sample $\{f_k|A_4\}$ as the \emph{prior}
and the data set $C_5$ as the new evidence. 
This setup aims to mimic the conditions at which NNPDF had studied 
the reweighting technique: the data set $A_4$ contains more data than 
$C_5$ and therefore the effects of including the later must be sub-dominant
for a global fit as well as the reweighting.
The results are shown in Fig.~\ref{fig:example5B}. 
It is clear that in this situation the reweighting of both methods
yield similar results compatible with global fits. 

One way to quantify the information about the parameters 
$\vec{\alpha}$ provided by the likelihood $p(x |\vec{\alpha})$, where $x$ is
either $\vec{y}$ or $\chi$ is to calculate the Kullback-Leibler (KL)  
divergence (see Appendix \ref{sec:kullbakc}.
Table \ref{tab:KL} shows the KL divergences for the reweighting 
results  performed above. 
The values in the table confirms the loss of information 
when using $\mathcal{P}(\vec{\alpha}|\chi)$ as the likelihood  instead of
$\mathcal{P}(\vec{\alpha}|\vec{y})$ in the reweighting procedure.

\begin{table}
\begin{center}
\begin{tabular}{|c|c||c|c||c|c|cc}\hline
SET & data & SET & data & SET & data \\ \hline
$A_0$ & $d_5$ &  &  &   &
\\\hline
$A_1$ & $d_4,d_5,d_6$ & $B_1$ & $d_4,d_6$ &  &
\\\hline 
$A_2$ & $d_3,d_4,d_5,d_6,d_7$ & $B_2$ & $d_3,d_4,d_6,d_7$ & &
\\ \hline 
$A_3$ & $d_2,d_3,d_4,d_5,d_6,d_7,d_8$ & $B_3$ & $d_2,d_3,d_4,d_6,d_7,d_8$ &&
\\\hline 
$A_4$ & $d_1,d_2,d_3,d_4,d_5,d_6,d_7,d_8,d_9$ & $B_4$ &
$d_1,d_2,d_3,d_4,d_6,d_7,d_8,d_9$ & &
\\\hline 
$A_5$ & $d_0,d_1,d_2,d_3,d_4,d_5,d_6,d_7,d_8,d_9,d_{10}$ & 
$B_5$ & $d_0,d_1,d_2,d_3,d_4,d_6,d_7,d_8,d_9,d_{10}$ & $C_5$ & 
$d_0,d_{10}$
\\\hline 
\end{tabular}
\end{center}
\caption{Data sets.}
\label{tab:datasplit}
\end{table}
\begin{table}
\begin{center}
\begin{tabular}{|c|c|c|c|c|c|cc}\hline
prior data & 
new evidence & 
$\mathcal{D}(\mathcal{P}(\vec{\alpha}|\vec{y})||\mathcal{P}(\vec{\alpha}))$&
$\mathcal{D}(\mathcal{P}(\vec{\alpha}|\chi)||\mathcal{P}(\vec{\alpha}))$
\\\hline
$A_0$&$B_1$&$2.94$&$1.86$\\\hline
$A_0$&$B_2$&$3.96$&$2.62$\\\hline
$A_0$&$B_3$&$4.77$&$3.13$\\\hline
$A_0$&$B_4$&$5.32$&$3.51$\\\hline
$A_0$&$B_5$&$5.84$&$3.94$\\\hline
$A_4$&$C_5$&$0.83$&$0.08$\\\hline
\end{tabular}
\end{center}
\caption{KL divergences.}
\label{tab:KL}
\end{table}
%

\section{Conclusions}
\label{sec:conclusions}

The technique of statistical inference is a useful tool to constrain
probability density functions in the presence of new evidence.
It is an alternative method to obtain updated distributions without having to 
perform a global fit by appending the old data and the new data.
The NNPDF collaboration has argued that the method proposed in 
~\cite{Ball:2010gb,Ball:2011gg}
is not adequate and they proposed their own method.
In the light of the results presented in this paper, we conclude that 
both methods are statistically equivalent in the limit when 
the prior densities are well constrained by the data and the new
evidence do not provide significant information.
We have shown using a numerical example that, if the uncertainties in 
the prior distribution are larger compared to the uncertainties obtained
by the inclusion of new data, the method proposed by NNPDF collaboration
is less efficient than the method proposed by 
~\cite{Giele:1998gw}
and the latter yields results that are significantly closer to those obtained 
from global fits.

\acknowledgments
We thank Seth Quackenbush for helpful discussions on the subject.
This work was supported by DOE contract No.~DE-SC0010102.

\appendix
\section{proof of Eq.~(\ref{eq:lh2})}
\label{sec:lh2}
The distribution $\mathcal{P}(\chi^2|\vec{\alpha})$ can be obtained by 
integrating $\mathcal{P}(\vec{y}|\vec{\alpha})$ subjected to 
$\chi^2=\chi^2(\vec{y},\vec{t})$. 
Mathematically this is simply  
\begin{align}
\mathcal{P}(\chi^2| \vec{\alpha})
=&	\int \delta[\chi^2 - \chi^2(\vec{y},\vec{t})] 
	\,p(y|\vec{\alpha})\,d^ny\notag\\
=&	\frac{1}{2\pi i} \int_{-\infty}^\infty d (i\omega) 
	\,e^{i\omega \chi^2} \int e^{-i \omega \chi^2(\vec{y},\vec{t})} 
	\,p(y|\vec{\alpha}) \,d^n y,\notag\\
=&	\frac{1}{2\pi i} \int_{-\infty}^\infty d (i\omega) \, 
	e^{i\omega \chi^2} \int \frac{1}{(2\pi)^{n/2}|\Sigma|^{1/2}} 
	e^{-\frac{1}{2} (2 i \omega + 1) \chi^2(y,\vec{\alpha})} 
	\,d^n y,\nonumber\\
=&  \frac{1}{2^{n/2}} \frac{1}{2\pi i} 
	\int_{-\infty}^\infty d (i\omega) \, e^{i\omega \chi^2}
	\frac{1}{(i \omega + 1/2)^{n/2}}, \nonumber\\
=&	\frac{1}{2^{n/2} \, \Gamma(n/2)} (\chi^2)^{\frac{1}{2}(n - 2)} 
	e^{-\frac{1}{2}\chi^2}.
\end{align}
Then we obtain 
\begin{align}
\mathcal{P}(\chi|\vec{\alpha})
=&	\int d\tilde{\chi}^2
	\,\delta[\chi-\tilde{\chi}] 
	\,\mathcal{P}(\tilde{\chi}^2|\vec{\alpha})\notag\\
=&	\frac{1}{2^{n/2-1} \, \Gamma(n/2)} (\chi^2)^{\frac{1}{2}(n - 1)} 
	e^{-\frac{1}{2}\chi^2}.
\end{align}
\section{the Hessian method}
\label{sec:hessian}
For completeness in this appendix we present the standard Hessian method
for error propagation. Suppose the model parameters $\vec{\alpha}_0$ 
that minimizes the $\chi^2$ is found. The method consists of expanding 
the $\chi^2$ around the minima as a function of the parameters:
\begin{align}
\chi^2(\vec{y},\vec{\alpha})
&\equiv
	\sum_{ij} (y_i-t_i(\vec{\alpha}))
	\Sigma^{-1}_{ij}
	(y_j-t_j(\vec{\alpha}))\notag\\
&\approx	
	\chi^2_0 + \sum_{ij} (\alpha_i-\alpha_i^0)C^{-1}_{ij}
	(\alpha_j-\alpha_j^0),
\label{eq:hessian}
\end{align}
where $C^{-1}_{ij}$ is the Hessian matrix given by
\begin{align}
H_{ij}=\frac{1}{2}\frac{\partial^2 \chi^2}{\partial a_i\partial a_j}
\end{align}
that is evaluated at  $\vec{\alpha}=\vec{\alpha}_0$. Next we diagonalize 
the matrix $C$ which gives eigenvectors $\vec{v}_j$ with eigenvalues 
$\lambda_j$. 
The displacements $(\vec{\alpha}-\vec{\alpha}_0)$ in Eq.~(\ref{eq:hessian})
can be written in terms of rescaled vectors $e_k=\sqrt{\lambda_k}\vec{v}_k$
\begin{align}
\delta \vec{\alpha}
\equiv	\vec{\alpha}-\vec{\alpha}_0
= \sum_k z_k \vec{e}_k
\label{eq:disp}
\end{align}
Replacing Eq.~(\ref{eq:disp}) in Eq.~(\ref{eq:hessian}) gives
\begin{align}
\chi^2(\vec{y},\vec{\alpha})
&\approx
	\chi^2_0 + \sum_{kq} z_k z_q (\vec{e}_k)^t C^{-1}_{ij}\vec{e}_q
	\notag\\
&=
	\chi^2_0 + \sum_{k} z_k^2 
\label{eq:chi2_z}
\end{align}
Notice that each displacements $\pm\delta\vec{\alpha}_k=\pm\vec{e}_k$
($z_k=1$) corresponds in Eq.~(\ref{eq:chi2_z}) a $\chi^2$ change of $1$ unit.
The interval defined by these displacements is known as the one-sigma 
confidence interval.
\section{Kullback-Leibler divergence}
\label{sec:kullbakc}
The Kullback-Leibler (KL) divergence~\cite{bib:KL} of the posterior density 
$\mathcal{P}(\vec{\alpha}|x)$ from the prior $\mathcal{P}(\vec{\alpha})$ 
is given by
\begin{align}
\mathcal{D}(\mathcal{P}(\vec{\alpha}|x) ||\mathcal{P}(\vec{\alpha})) 
&=	\int\mathcal{P}(\vec{\alpha} | x) \,
	\ln \frac{\mathcal{P}(\vec{\alpha}|x)}{\mathcal{P}(\vec{\alpha})}\,
	 d^n\alpha, \nonumber\\
&=	\int \frac{\mathcal{P}(x|\vec{\alpha})}{\mathcal{P}(x)} \, 
	\mathcal{P}(\vec{\alpha}) \, 
	\ln \frac{\mathcal{P}(x|\vec{\alpha})}{\mathcal{P}(x)} \,  
	d^n\alpha, \nonumber\\
&\approx
	\frac{1}{N} \sum_{k=1}^N w_k \, \ln w_k,
	\label{eq:KL}
\end{align}
where the weights are defined as in Sec.~\ref{sec:the reweighting method}.
The larger the  KL divergence, the greater the difference between 
$\mathcal{P}(\vec{\alpha}| x)$ and $\mathcal{P}(\vec{\alpha})$ and, 
therefore, the  more informative are the data $x$ about the PDF parameters, 
relative to what was known about them prior to inclusion of these data.
A similar quantity called \emph{effective} number of replicas 
$N_{\text{eff}}$ 
was defined in the references \cite{Ball:2010gb,Ball:2011gg}:
\begin{align}
N_{\text{eff}}
=\exp\left(\frac{1}{N}\sum_kw_k \ln\left(\frac{N}{w_k}\right)\right)
\end{align}
Here $N$ is the number of Monte Carlo sample (\emph{replicas}) taken
from the prior distribution.
Clearly the KL divergence is related to $N_{\text{eff}}$ via
\begin{align}
\mathcal{D}(\mathcal{P}(\vec{\alpha}|x) ||\mathcal{P}(\vec{\alpha})) 
\approx \ln\left(\frac{N_{\text{eff}}}{N}\right)
\end{align}


\begin{thebibliography}{99}

\bibitem{Giele:1998gw} 
  W.~T.~Giele and S.~Keller,
  Phys.\ Rev.\ D {\bf 58}, 094023 (1998)
  [hep-ph/9803393].


\bibitem{Ball:2010gb} 
  R.~D.~Ball {\it et al.}  [NNPDF Collaboration],
  Nucl.\ Phys.\ B {\bf 849}, 112 (2011)
  [Erratum-ibid.\ B {\bf 854}, 926 (2012)]
  [Erratum-ibid.\ B {\bf 855}, 927 (2012)]
  [arXiv:1012.0836 [hep-ph]].


\bibitem{Ball:2011gg} 
  R.~D.~Ball, V.~Bertone, F.~Cerutti, L.~Del Debbio, S.~Forte, A.~Guffanti, N.~P.~Hartland and J.~I.~Latorre {\it et al.},
  Nucl.\ Phys.\ B {\bf 855}, 608 (2012)
  [arXiv:1108.1758 [hep-ph]].





\bibitem{bib:Bernardo} 
J.M.~Bernardo, \url{http://www.uv.es/~bernardo/teaching.html}


\bibitem{bib:Gillespie} 
D.T.~Gillespie,
``A theorem for physicists in the theory of random variables",
Am. J. Phys. {\bf 51}, 520 (1983).

\bibitem{bib:KL}
\emph{Ibid.}~\cite{bib:Bernardo}. Note, this reference uses the notation $\kappa(q | p) = D(p || q)$.


\end{thebibliography}
\end{document}